\documentclass{ws-procs975x65}

\begin{document}

\title{Quark-Lepton Symmetry and Quartification in Five
  Dimensions\footnote{Talk given at the Festschrift in honour of Bruce
    McKellar and Girish Joshi at the University of Melbourne, 14/12/2006. To
    appear in a special edition of IJMPA. The talk is based on the results
  of
  references~\cite{McDonald:2006dy,Coulthurst:2006bc,Coulthurst:2006kz,Demaria:2006bd}~.
  }}

\author{Kristian L. McDonald$^\dagger$}

\address{Research Center for High Energy Physics, University of Melbourne,\\
Victoria, 3010, Australia\\
$^\dagger$E-mail: k.mcdonald@physics.unimelb.edu.au\\}

\begin{abstract}
We outline some features of
higher dimensional models possessing a Quark-Lepton (QL) symmetry. The
QL symmetric
model in five dimensions is discussed, with particular emphasis on the
use of split
fermions. An interesting fermionic geography which utilises the QL
symmetry to suppress the proton decay rate and to motivate the flavor
differences in the quark and leptonic sectors is considered. We
discuss the quartification model in five dimensions and contrast the features
of this model with traditional four dimensional constructs.
\end{abstract}


\bodymatter
\section{Introduction}\label{sec_intro}
The Standard Model (SM) of particle physics displays a clear asymmetry
between quarks and leptons. Quarks and leptons have different masses
and charges and importantly quarks experience the strong
force. Despite these differences a suggestive
similarity exists between the family structure of quarks and leptons, leading
one to wonder if the SM may be an approximation to a
more symmetric fundamental theory.

The similar family structure of quarks and leptons is an automatic
consequence of the defining symmetry of the Quark-Lepton (QL)
symmetric model~\cite{Foot:1990dw}. In this framework the similarity
between quarks and leptons is elevated to an exact symmetry of
nature. The model
permits a complete interchange symmetry between quarks and a
generalized set of leptons, with the SM
resulting from the breaking of an enlarged symmetry group.

On an independent front, a number of model building
tools which employ extra spatial dimensions have been developed in
recent years. In particular new methods of symmetry breaking have
been uncovered. These methods allow one to reduce the scalar content required
to achieve symmetry breaking in four dimensionsal models. It is
natural to investigate these methods within pre-existing frameworks to
determine the phenomenological distinctions between the new and
traditional approaches.

In this work we
outline some recent investigations in five dimensional
models possessing a QL symmetry. We focus our attention on two aspects
of these investigations. First we consider the use of split fermions
in models with a QL symmetry. We show that the symmetry constrains the
implementation of split fermions and allows one to solve some
outstanding problems in four dimensional with QL models; namely how can
one resolve the notion of a QL symmetry with the disimilarity
between the observed masses of quark and lepton.

We also consider the
quartification model, a framework which extends the notion
of a QL symmetry to permit gauge
unification, in five dimensions. We show that the higher dimensional
quartification model
allows one to remove many of the complications which arise in four
dimensional models. These complications result mainly from the relatively
large Higgs sector
required to achieve the necessary symmetry breaking. We show that an
effective Higgsless limit may be obtained in the five dimensional
quartification model, thereby permitting considerable simplification.    
\section{Quark-Lepton Symmetry}\label{sec:defining_ql}
How does one construct a
quark-lepton symmetric model? Let us recall that the SM
also displays a clear asymmetry between left and right handed fields;
the left handed fields
experience $SU(2)_L$ interactions whilst the right handed fields do
not. One generation of SM fermions
may be denoted as
\begin{eqnarray}
Q,L,u_R,d_R,e_R,
\end{eqnarray}
revealing a further left-right asymmetry; namely the absence of
$\nu_R$. However the left-right asymmetry may be a purely low energy
phenomenon and the construction of a high energy left-right symmetric
theory proceeds
as follows. One must first extend the SM
particle content to include $\nu_R$ and thereby equate the number of
left and right degrees of freedom. The SM gauge
group must also be enlarged~\cite{Pati:1974yy,Mohapatra:1974hk,Mohapatra:1974gc,Senjanovic:1975rk}:
\begin{eqnarray}
SU(3)\otimes SU(2)\otimes U(1)\rightarrow SU(3)\otimes
[SU(2)]^2\otimes U(1),\nonumber
\end{eqnarray}
where the second $SU(2)$ group acts on right-chiral fermion doublets. This enables one to define a discrete symmetry interchanging all left
and right handed fermions in the Lagrangian,
$f_L\leftrightarrow f_R$. Furthermore the model must be constructed
such that
the additional symmetries introduced to enable the
left-right interchange
are suitably broken to reproduce the SM at low energies. In practise
this means that a suitable extension of the Higgs sector is also required.

One may construct a quark-lepton symmetric model by employing the same
recipe~\cite{Foot:1990dw,Foot:1990um,Foot:1990un,Foot:1991fk,Levin:1993sq,Shaw:1994zs,Foot:1995xx}.
First  the fermion content of the SM must be extended. As quarks
come in three colors one is required to introduce more leptons to
equate the number of quark and lepton degrees of freedom. For each SM
lepton one includes two exotic leptons,
\begin{eqnarray}
e\rightarrow E=(e,e',e''),\nonumber\\
\nu\rightarrow N=(\nu, \nu',\nu''),
\end{eqnarray}
where the primed states are the exotics (known
as liptons in the literature). The gauge group must also be extended:
\begin{eqnarray}
SU(3)\otimes SU(2)\otimes U(1)\rightarrow [SU(3)]^2\otimes SU(2)\otimes U(1),\nonumber
\end{eqnarray}
where the additional $SU(3)$ gauge bosons induce transitions amongst
the generalized leptons in the same way that gluon exchange enables quarks to
change color. Now one may define a discrete symmetry
interchanging all quarks and (generalized) leptons in the
Lagrangian,
\begin{eqnarray}
Q\leftrightarrow L,\mkern15mu u_R\leftrightarrow E_R,\mkern15mu
d_R\leftrightarrow N_R.
\end{eqnarray}
The  gauge symmetry may be broken via the Higgs mechanism to reproduce the
charge differences between quarks and leptons and give heavy masses
to the liptons. Phenomenologically consistent models can be obtained
by breaking the lepton color group $SU(3)_\ell$ completely or by
leaving an unbroken subgroup $SU(2)_\ell\subset SU(3)_\ell$ (which
serves to confine the liptons into exotic bound states). The QL
symmetry implies mass
relations of the type $m_u=m_e$ which are more difficult to rectify. In 4D
one may remove these by extending the scalar sector and thereby
increasing the number of independent Yukawa couplings.
\section{QL Symmetry in Five Dimensions}\label{sec:ql_5d}
In recent years the study of models with extra dimensions has revealed
a number of new model building tools. The use of orbifolds provides a
new means of reducing the gauge symmetry operative at low energies by
introducing a new mass scale, namely the compactification scale
$M_c=1/R$. In generic 5D models the mass of exotic gauge bosons can be
set by $M_c$ and the
reduction of symmetries by orbifold construction results in collider
phenomenology which differs from that obtained in models employing the
usual 4D Higgs symmetry
breaking.

The reduction of the QL symmetric gauge group via orbifold
construction has recently been studied in
5D~\cite{McDonald:2006dy}. One question that faces the model builder
in extra dimensional models is whether to place fermions on a brane or
in the bulk. We shall focus on the latter in what follows and
demonstrate that the troublesome mass relations which arise in 4D QL
models actually provide useful and interesting model building
constraints in 5D models~\cite{Coulthurst:2006zu}. We note that
placing fermions in
the bulk permits the unification of quark and lepton
masses at TeV scales in more a general class of
models~\cite{Hung:2004ac,Adibzadeh:2007hz}. 

In five dimensions bulk fermions lack chirality. However, chiral zero
mode fermions, which one may identify with SM fermions, may be
obtained by employing orbifold boundary conditions on a fifth
dimension forming an $S^1/Z_2$ orbifold. By coupling a
bulk
fermion to a bulk scalar field one may readily localise a chiral
zero mode fermion at one of the orbifold fixed
points~\cite{Georgi:2000wb}. Denoting the bulk fermion (scalar) as
$\psi$ ($\phi$) one has the following Lagrangian:
\begin{eqnarray}
\mathcal{L}=\bar{\psi}(\Gamma_M \partial^M -f\phi)\psi
-\frac{1}{2}\partial_M\phi\partial^M\phi -V(\phi),
\end{eqnarray}
where $\Gamma_M$ are the Dirac matrices, $f$ is a constant,
$M=0,1...4$ is the 5D Lorentz index and
\begin{eqnarray}
V(\phi)=\frac{\lambda}{4}(\phi-v)^2,
\end{eqnarray}
is the usual quartic potential. If $\phi$ transforms trivially under
$Z_2$ its ground state is given by $\langle\phi\rangle=v$. However if
$\phi$ is odd under $Z_2$ its ground state is required to vanish at
the fixed points. This results in a kink vacuum profile for $\phi$
which serves to
localise chiral zero mode fermions, $\psi_0$, at one of the orbifold fixed
points. The fixed point at which $\psi_0$ is localised depends on the
sign of $f$.

\begin{figure}[b] 
\centering
\includegraphics[width=0.6\textwidth]{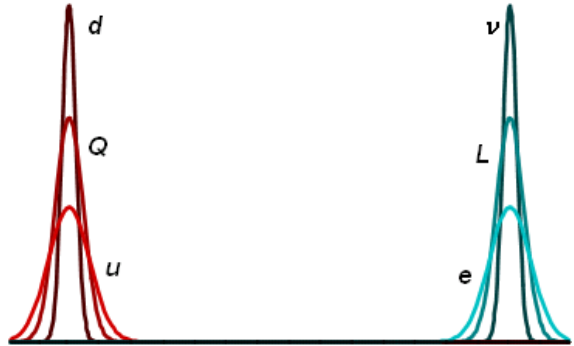} 
  \caption{The 5D wavefunctions for quarks and leptons in a QL
    symmetric model with one bulk scalar.}
  \label{fig:talk_jmp_one_scalar}
\end{figure}

In a QL symmetric model one must specify the transformation properties
of the bulk scalar under the QL symmetry. An interesting choice is to
make $\phi$ odd under the QL symmetry~\cite{Coulthurst:2006bc}, resulting in a Yukawa Lagrangian of
the form
\begin{eqnarray}
\mathcal{L}=-\left\{f_Q(Q^2-L^2) +f_u(U^2 -E^2) +f_d(D^2-N^2)\right\}\phi,\nonumber
\end{eqnarray}
where $Q^2=\bar{Q}Q$, etc, and the SM fermions are identified with
the chiral zero modes of the bulk fermions in an obvious fashion. Note
that the choice $f_Q,f_u,f_d>0$ automatically implies the localisation
pattern shown in Figure~\ref{fig:talk_jmp_one_scalar}.
It is of interest that this geography may be implemented in a less
arbitrary fashion in a QL symmetric framework as this pattern is
precisely that advocated recently to suppress
the proton decay rate without an extremely large
ultra-violet cutoff~\cite{Arkani-Hamed:1999dc}. Indeed the effective 4D proton decay inducing
non-renormalizable operator has the form
\begin{eqnarray}
\mathcal{O}_p=\frac{K}{\Lambda^2}\mathcal{O}_Q^3\mathcal{O}_L 
\end{eqnarray}
where $\mathcal{O}_Q$ ($\mathcal{O}_L$) generically denotes a quark
(lepton) operator and $K\sim \exp\left\{-vL^{3/2}\right\}$ represents the wavefunction overlap between
quarks and leptons in the extra dimension. $L$ denotes the length
of the fundamental domain of the orbifold.

Observe that fermions related by the QL symmetry necessarily develop
identical wavefunction profiles
in the extra dimension and consequently the troublesome mass relations
of the type $m_e=m_u$ persist in the effective 4D theory. With one
bulk scalar it is only possible to localise fermions at the orbifold
fixed points. However two bulk scalar models enable one to localise
fermions within the bulk. This works as
follows~\cite{Grossman:2002pb,Grossman:2003sr}. With one bulk scalar
chiral zero mode fermions are found at one of the orbifold fixed
points, with the precise point of localisation determined by the sign
of the relevant Yukawa coupling. If a second bulk scalar is added with
an opposite sign Yukawa coupling it will tend to drag the fermion
towards the other end of the extra dimension, thereby localising it in
the bulk. Importantly though the chiral fermion cannot be pulled very
far into the bulk before it is dragged all the way to the other end of
the extra dimension. 

\begin{figure}[b] 
\centering
\includegraphics[width=0.6\textwidth]{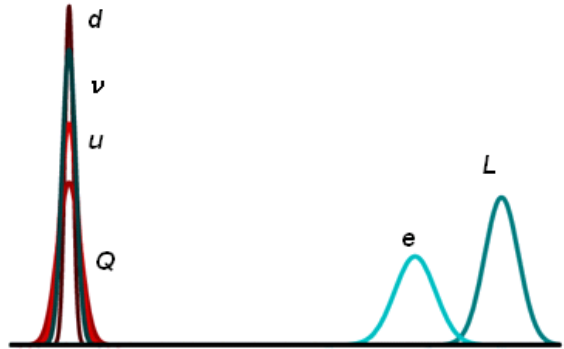} 
  \caption{The 5D wavefunctions for quarks and leptons in a QL
    symmetric model with two bulk scalars.}
  \label{fig:talk_jmp_two_scalars}
\end{figure}

Let us add a second bulk scalar which is even under the QL symmetry,
giving rise to the Lagrangian
\begin{eqnarray}
\mathcal{L}_2&=&-\left\{h_Q(Q^2+L^2) +h_u(U^2 +E^2)\right.\nonumber\\
& &\left.\mkern45mu +h_d(D^2+N^2)\right\}\phi'.\nonumber
\end{eqnarray}
Let us again require $h_Q,h_u,h_d>0$ so that all quark Yukawa
couplings are positive. Both $\phi$ and $\phi'$ will attempt to
localize quarks at the same point in the extra dimension and they
will remain at their original point of localisation. This type of
quark geography is precisely that recently advocated~\cite{Lillie:2003sq} to
allow one
to construct quark flavor without introducing large flavor changing
neutral currents. However $\phi$
and $\phi'$ attempt to localise leptons at different fixed points and
the resultant point of localisation for a given lepton depends on
which scalar dominates (these statements will be made numerically
precise in a forthcoming publication~\cite{5D_ql_flavour_study}). An
arrangement typical of this setup is shown in
Figure~\ref{fig:talk_jmp_two_scalars}.

This arrangement has a number of interesting features. Firstly note that
shifting leptons into the bulk significantly alters the degree of
wavefunction overlap in the quark and leptonic sectors. When one
obtains the effective 4D fermion masses the different degree of
overlap in the quark and lepton sectors will remove the undesirable QL
mass relations. Furthermore the overlap between left and right chiral
fermion wavefunctions is expected to be greater in the quark sector
than in the lepton sector, leading to the generic expectation that
quarks will be heavier than the lepton to which they're related by the
QL symmetry. Observe that dragging the right chiral neutrinos $\nu_R$
all the way to the `quark
end' of the extra dimension significantly suppresses the neutrino Dirac
masses below the electroweak scale. Dragging $\nu_R$ to the quark end
of the extra dimension does not introduce rapid proton decay, provided
that $\nu_R$ has a large enough Majorana mass to kinematically
preclude decays of the type $p\rightarrow \pi \nu_R$. Such a mass is
expected to arise at the non-renormalizable level even if it is not
induced by tree-level couplings in the
theory~\cite{Coulthurst:2006kz}.

Thus the non-desirable Yukawa relations implied by 4D QL symmetric
models turn out to be of interest in the 5D construct. They enable one
to understand proton longevity within the split fermion framework in a
less arbitrary fashion and instead of inducing unwanted mass relations
suggest an underlying motivation for the flavour differences
experimentally observed
in the quark and lepton sectors.
\section{Quartification in Five Dimensions}\label{sec:quart}
Adding a leptonic color group to the SM clearly renders the
traditional approaches to gauge unification inapplicable. Recall that the simplest grand unified theory, namely $SU(5)$,
does not contain the left-right symmetric model. However there are
larger unifying groups which do admit the left-right symmetry, for
example $SO(10)$ and the trinification gauge group $[SU(3)]^3\times
Z_3$. Similarly it is possible to construct a unified
gauge theory admitting the QL symmetry by considering larger
unification groups. It is natural to extend the notion of
trinification to include a leptonic color factor, leading one
to the so called quartification model. This possesses
the gauge group
$G_Q=[SU(3)]^4\times Z_4$, where the additional $SU(3)$ factor
corresponds to lepton color and the $Z_4$ symmetry cyclicly permutes
the group factors to ensure a single coupling constant~\cite{Joshi:1991yn,Babu:2003nw,Chen:2004jz,Demaria:2005gk,Demaria:2006uu,Demaria:2006bd}.

It was been demonstrated that unification may be achieved within the
quartification framework in 4D by enforcing additional
symmetries upon the quartification model~\cite{Babu:2003nw}. Subsequent
work has shown that unification need not require additional
symmetries, but does require multiple symmetry breaking scales between
the unification scale and the electroweak
scale~\cite{Demaria:2005gk}. It was also shown~\cite{Demaria:2005gk}
that unification may be
achieved via multiple symmetry breaking routes both with and without
the remnant leptonic color symmetry $SU(2)_\ell$. 

The necessary
symmetry breaking is accomplished with eight Higgs multiplets, giving
rise to a complicated Higgs potential with a large number of free
parameters. The demand of multiple symmetry breaking scales also requires
hierarchies of vacuum expectation values (VEV's) to exist within individual
scalar multiplets, giving rise to a generalized version of the
doublet-triplet splitting problem familiar from $SU(5)$ unified
theories. A large number of electroweak doublets also appear in the 4D
constructs. Thus many of the less satisfactory features of 4D
quartification models revolve around the Higgs sector required for symmetry
breaking.

Recently the quartification model has been studied in 5D, where
intrinsically higher dimensional
symmetry breaking methods exist~\cite{Demaria:2006bd}. By
taking the fifth dimension as an $S^1/Z_2\times Z_2'$ orbifold one
may employ 
orbifold boundary conditions (OBC's) on the bulk gauge sector to reduce the
gauge symmetry operative at the zero mode level from $G_Q$
to
\begin{eqnarray}
SU(3)_c\otimes SU(2)_L\otimes SU(2)_\ell\otimes SU(2)_R\otimes U(1)^3.
\end{eqnarray}
However the use of OBC's does not reduce the rank of the gauge group
so that further symmetry breaking is required. Rank reducing symmetry breaking
can be achieved in higher dimensional theories by employing a boundary
scalar sector to alter the boundary conditions on the compactified
space for gauge fields~\cite{Csaki:2003dt}. Denoting a boundary scalar
as $\chi$ and defining $V\propto \langle\chi\rangle$ one can show that
$V$ induces a shift in the Kaluza-Klein mass spectrum of
gauge fields which couple to $\chi$. If such a gauge field
initially possessed a massless mode its tower receives a shift of the form
\begin{equation}\label{eq:shift1}
\centering
M_n \approx M_c \,(2n+1) \left(1+\frac{M_c}{\pi V}+ \dots \right),
\mkern15mun=0,1,2,...
\end{equation}
giving a tower with the lowest-lying states $M_c, \, 3M_c, \, 5M_c, ...$. This represents an offset of 
$M_c$ relative to the $V=0$ tower, with the field no longer retaining
a massless zero mode. The association of $V$ with the VEVs of the
boundary scalar
sector implies that the limit $V \rightarrow \infty$ is attained when $\langle \chi \rangle \rightarrow \infty$.
However, when the VEVs of the Higgs fields are taken to infinity, the shift in the KK masses of the gauge fields is finite, 
giving the exotic gauge fields masses dependent only upon the compactification scale $M_c$. 
Consequently, these fields remain as ingredients in the effective theory while the boundary Higgs sector decouples entirely, and we 
can view our reduced symmetry theory in an effective Higgsless limit. Interestingly, in this limit also, the high-energy 
behaviour of the massive gauge boson scattering remains unspoilt as
shown in~\cite{Csaki:2003dt}.

It was shown in~\cite{Demaria:2006bd} that a unique set of OBC's was required
to ensure that quark masses could be generated and to prevent liptons
from appearing at the electroweak scale. The inclusion of a boundary
Higgs sector allows one to
reduce the
quartification gauge symmetry down to the SM gauge group $G_{SM}$ or to
$G_{SM}\otimes
SU(2)_\ell$ at the zero mode level. In both cases fifth
dimensional components of the $SU(3)_L$ gauge fields with the quantum
numbers of the SM Higgs doublet retained a massless mode, enabling one
to use Wilson loops to reproduce the SM flavour structure.

A surprising result however was that unification could only be
achieved when the remnant lepton color symmetry $SU(2)_\ell$ remained
unbroken. Thus one arrives at a unique minimal quartification model
which unifies in 5D, a result to be contrasted with the 4D case where a
large number of symmetry breaking routes which permit unification have
been uncovered. Unfortunately the unifying case requires
the compactification scale to be greater than $10^{10}$~GeV so that only a
SM like Higgs field is expected to appear at the LHC.
\section{Conclusion}\label{sec:conclusion}
Quark-lepton symmetric models in some sense unify the fermionic
content of the SM and thereby 
motivate the similar family structures observed in
the quark and lepton sectors. Recent investigations involving QL symmetries
in 5D have uncovered a number of interesting results. In particular the QL symmetry provides useful Yukawa relationships in split fermion
models and the quartification model is found to be more constrained in 5D. A
number of avenues for further investigation remain, including a detailed
analysis of the fermionic geography of
Figure~\ref{fig:talk_jmp_two_scalars}~\cite{5D_ql_flavour_study} and
the construction of a complete three generational model with combined
QL and left-right symmetry in 5D~\cite{Coulthurst:2007rj}.

\end{document}